\documentclass[twocolumn,showpacs,amsmath,amssymb,aps,pra,floatfix,nofootinbib,10pt]{revtex4-2}
\usepackage[latin1]{inputenc}
\usepackage[english]{babel}
\usepackage[T1]{fontenc}
\usepackage{mathrsfs}
\usepackage{hyperref}
\usepackage{multirow}
\usepackage{bm,epsfig}
\usepackage{graphicx}
\usepackage{subeqnarray}
\usepackage{bbold}
\usepackage{upgreek}
\hypersetup{colorlinks=true, urlcolor=blue, citecolor=blue, filecolor=blue, linkcolor=blue}
\usepackage{dcolumn}
\usepackage{amsmath}
\usepackage{color}
\usepackage{siunitx}
\usepackage{physics}

\usepackage{xcolor}

\renewcommand{\Re}{\operatorname{Re}}

\def\beq#1{\begin{equation}\label{#1}}
\def\eeq{\end{equation}}

\begin{document}

\title{%
A simple analytical alignment model for laser-kicked molecular rotors
}

\author{
A.~L\"ohr$^{1,2,\ast}$, M.~Ivanov$^{1,2,3}$ and M.~Khokhlova$^{1,4}$
}
\affiliation{
\mbox{$^{1}$Max Born Institute for Nonlinear Optics and Short Pulse Spectroscopy, 
12489 Berlin, Germany} \\
\mbox{$^{2}$Department of Physics, Humboldt University, 
12489 Berlin, Germany} \\
\mbox{$^{3}$Blackett Laboratory, Imperial College London, SW7 2AZ London, UK}\\
\mbox{$^{4}$Department of Physics, King's College London, WC2R 2LS London, UK} \\
\mbox{$^{\ast}$alexander.loehr@mbi-berlin.de}
}

\begin{abstract}
We develop a mathematically simple yet accurate model for the single-pulse non-resonant impulsive alignment of thermal ensembles of linear molecules. We find that our molecular alignment model not only provides a simplification for analytical and numerical calculations, but also establishes intuitive connections between system parameters, such as gas temperature and laser pulse intensity, and the resulting shape of the temporal molecular alignment.
\end{abstract}

\maketitle
\noindent

\section{Introduction}
Non-resonant laser-induced alignment is ubiquitous in virtually all experiments and processes that concern the interaction of strong polarized light with low-density media containing linear molecules, such as attosecond spectroscopy~\cite{bruner2016multidimensional, niikura2003probing}, laser-induced electron diffraction~\cite{spanner2004reading}, laser filamentation and air lasing~\cite{kartashov2014transient, britton2019short, lytova2020n, richter2020rotational}. Considerable analytical and numerical efforts have been conducted to describe the alignment process, particularly in the non-adiabatic, short-pulse regime~\cite{Leibscher2004, ghafur2009impulsive, broege2008strong, owschimikow2011laser, zhao2003alignment}.

Computing microscopic alignment dynamics for any given set of laser parameters numerically is relatively straightforward~\cite{fleischer2008selective, torres2005dynamics, renard2005control}. Nevertheless, a simple analytical model would always be preferred, especially if one has to perform microscopic alignment calculations for many sets of laser parameters. This situation arises, for example, in the case of macroscopic propagation of an intense laser pulse through a dense molecular gas with inevitable reshaping of the propagating pulse along the~way. 

Analytical results, like those of the three-dimensional~(3D) kicked-rotor model~\cite{Leibscher2004}, offer some relief, but still require the numerical execution of numerous infinite sums and complex hypergeometric functions. A simple and intuitive analytical model of the alignment process, while desired, is still lacking. 

We develop such a simplified alignment model in this paper, which is structured as follows. We begin by laying the groundwork for our theoretical treatment of the kicked quantum rotor in Sec.~\ref{SectionGeneralTheory} by defining the system Hamiltonian in terms of dimensionless parameters and derive the phase shift of the rotor wavefunction induced by a short laser kick.

We first derive an analytical solution for the time-dependent alignment of a thermal ensemble of 2D delta-kicked rotors in Sec.~\ref{Section2D}, which is mathematically vastly simpler then its 3D counterpart. Encouraged by the similarity of the 2D and 3D alignment at higher temperatures, we apply physically-motivated modifications to the 2D alignment function in Sec.~\ref{SectionHR} and obtain a simple, yet powerful approximation for the alignment of a thermal ensemble of kicked 3D rotors. In Sec.~\ref{SectionMR} we simplify this approximate alignment model for quantum rotors even further, allowing us to extend it to linear molecules with arbitrary nuclear spin statistics and, therefore, to include fractional rotational revivals as well.

We find that for a wide range of parameters the temporal alignment of an initially thermal distribution of linear molecules can be approximated very well by a train of Gaussian envelopes each multiplied by an individual sinusoidal carrier. In Sec.~\ref{SectionResult} we demonstrate the performance of the alignment model and gauge it against the exact analytical 3D alignment model~\cite{Leibscher2004}, finding excellent agreement for a large range of system parameters. Finally, we present the alignment of real atmospheric molecules kicked by the same laser pulse.

\section{The kicked rotor}\label{SectionGeneralTheory}
We start from closely following the analytical approach~\cite{Leibscher2004}, which describes the 3D rigid rotor kicked by a short laser pulse. We assume that individual rotors do not interact with each other, therefore the behavior of the system can be understood through a single molecule model. A quantum-mechanical molecule in an arbitrary state can be described as a superposition of its eigenstates. Ignoring translational movement and assuming that the molecule is rigid, e.g. there is no intramolecular movement, the wavefunction only has to account for rotational dynamics:
\begin{align}
 \ket{\Psi(\vec{r},t)} = \ket{\Psi(\phi,\theta,t)} \, .
\end{align}

Thus, the resulting Hamiltonian contains rotational kinetic energy and external angular potential:
\begin{align}
    \hat{H}(\phi,\theta,t)=\frac{\hat{L}^2}{2\mathcal{I}}+\hat{V}(\phi,\theta, t) \, ,
\label{ham_initial}
\end{align}
where $\mathcal{I}$ is the moment of inertia of the rotor, $\hat{L}$ is the angular momentum operator and $\hat{V}$ is the potential operator. 
In case of a linearly-polarized pump, the interaction can be described by the cycle-averaged angular potential 
\begin{align}
    \hat{V}(t,\theta)= -\frac{E_0^2(t) }{4} \left[ (\alpha_\parallel-\alpha_\perp)\cos^2(\theta) + \alpha_\perp\right] \, ,
\end{align}
where $\alpha_{\parallel}$ and $\alpha_{\perp}$ are the parallel and perpendicular polarizability of the molecule, respectively, and $E_0$ is the time-dependent amplitude of the pump field.

By introducing the dimensionless time $\tau=t\hbar/\mathcal{I}$ and the interaction strength $\epsilon(\tau)=(\alpha_\parallel-\alpha_\perp)E_0^2(\tau)\mathcal{I}/(4\hbar^2)$, the system Hamiltonian~\eqref{ham_initial} simplifies to the form
\begin{align}
    \hat{H}(\theta,\tau)=\frac{\hat{L}^2}{2}-\epsilon(\tau)\cos^2(\theta) \, .
\end{align}
Note that the angle-invariant term of the potential is omitted, as it does not cause any transitions between rotational states.

For a pump pulse duration significantly shorter than the rotational period of the rotor, $t \ll 2\pi \mathcal{I}/\hbar$, the so-called impulsive alignment regime, the pulse envelope can be approximated as a delta peak. For a pulse centered at $\tau=0$ the resulting Hamiltonian takes the form 
\begin{align}
     \hat{H}(\theta,\tau)=\frac{\hat{L}^2}{2}-\delta(\tau)\mathrm{P} \cos^2(\theta) \, ,
    \label{Hamiltonian}
\end{align}
where the integrated interaction strength $\mathrm{P}$ is defined as
\begin{align}
    \mathrm{P}=\int_{-\infty}^\infty \epsilon(\tau) d\tau = \frac{(\alpha_\parallel-\alpha_\perp)}{4\hbar} \int_{-\infty}^\infty E_0^2(t) dt \, .
    \label{defP}
\end{align}

We integrate the Schr\"odinger equation with the Hamiltonian~\eqref{Hamiltonian}:
\begin{align}
    -i\frac{\partial}{\partial \tau}\ket{\Psi(\tau)}= \frac{1}{2}\hat{L}^2\ket{\Psi(\tau)} - \delta(\tau)\mathrm{P} \cos^2(\theta)\ket{\Psi(\tau)} \, ,
\end{align}
which is equivalent to 
\begin{equation}
    -i\int_{\ket{\Psi(-\Delta)}}^{\ket{\Psi(\Delta)}} \frac{d\ket{\Psi(\tau)}}{\ket{\Psi(\tau)}} =  - \mathrm{P} \cos^2(\theta) + \int_{-\Delta}^{\Delta} \frac{\hat{L}^2\ket{\Psi(\tau)}}{2\ket{\Psi(\tau)}} d\tau
\end{equation}
in the limit $\Delta\xrightarrow{}0$. The right hand integral vanishes in the limit as the eigenvalues of the $\hat{L}^2$ are finite valued. Applying the limit yields the instantaneous response of the wavefunction to the delta pulse as
\begin{align}
   \ket{\Psi(0^+)}=\ket{\Psi(0^-)}e^{i \mathrm{P} \cos^2(\theta)} \, .
  \label{wf_evol}
\end{align}

The resulting time evolution of the wavefunction~\eqref{wf_evol} and its corresponding alignment have been fully solved analytically by Leibscher with co-workers~\cite{Leibscher2004}. The final alignment equation for the 3D case~\eqref{3DAlignment}, displayed in Appendix~\ref{AppendixLeibscher}, is a lengthy expression containing several infinite sums, Clebsch-Gordan coefficients and hypergeometric functions.

\section{2D rotor alignment}\label{Section2D}
To find a simplified solution to the 3D alignment problem, we turn to the case of a 2D rotor, which is mathematically far less complex and way more intuitive than the 3D one. For the calculation of the temporal alignment of the kicked 2D rotor, we follow in close analogy the 3D derivation~\cite{Leibscher2004}. 

In 2D the position of the rotor can be described as a point on a circle. As such, the angular eigenstates of the 2D rigid rotor are characterized by a single quantum number, and we can represent an arbitrary wavefunction as a superposition
\begin{align}
    \Psi(\theta,\tau)= \sum_{l=-\infty}^\infty c_l(\tau) \ket{l} \, ,
    \label{2DeigenstatesSum}
\end{align}
where 
\begin{align}
    \ket{l}=\sqrt{\frac{1}{2\pi}}e^{i l \theta}
    \label{2Dbasis}
\end{align}
is the eigenstate corresponding to the orbital quantum number $l$ and $c_l(\tau)$ is the projection of $\Psi(\theta,\tau)$ onto the state $\ket{l}$. In absence of an external potential the individual states evolve in time according to their rotational energies and are described by their time-dependent complex amplitudes as
\begin{align}
      c_{l}(\tau) = c_{l}  e^{-i\frac{l^2}{2}\tau} \, .
      \label{2DStatepropagation}
\end{align}

As eigenstates with different quantum numbers are orthogonal and eigenstate multiplication is trivial in the 2D basis, see Eq.~\eqref{2Dbasis}, it is useful to project the exponential factor in Eq.~\eqref{wf_evol} onto the basis states:
\begin{align}
     e^{i\mathrm{P}\cos^2(\theta)} =  \sum_{l=-\infty}^{\infty} d_{l} \ket{l}
     \label{expansion1}
    \, ,
\end{align}
where $d_l$ are the projection coefficients. From Eq.~\eqref{expansion1} using the orthogonality, the coefficients can be found as
\begin{align}
    d_l=
    \sqrt{\frac{1}{2\pi}}e^{i\frac{\mathrm{P}}{2}} \int_0^{2\pi} e^{i\frac{\mathrm{P}}{2}\cos(2\phi)} e^{i l \phi} d\phi \, ,
\end{align}
where the integral vanishes for odd values of $l$, and for even values equals
\begin{align}
    d_{2l}=  \sqrt{2\pi} i^l e^{i\frac{\mathrm{P}}{2}}  J_l\left(\frac{\mathrm{P}}{2}\right) \, .
    \label{d2coeff}
\end{align}

We can use Eq.~\eqref{d2coeff} to express the effect of the pump-induced kick on an arbitrary initial eigenstate according to Eq.~\eqref{wf_evol}. Assuming $\ket{\Psi(0^-)} = \ket{l_0}$, we can write
\begin{align}
    e^{i \mathrm{P} \cos^2(\theta)} \ket{l_0} = \sum_{l=-\infty}^\infty d_{2l} \sqrt{\frac{1}{2\pi}} \ket{l_0+2l} \, .
    \label{siglestateinteraction}
\end{align}
Furthermore, we can express the temporal alignment of the resulting rotational wavepacket, using the fact that the kicked system evolves accordingly to the energies of its constituent eigenstates, Eq.~\eqref{2DStatepropagation}.

We define the alignment as the expectation value of $\cos^2(\theta)$:
\begin{equation}
    \begin{split}
    \langle \cos^2(\theta) \rangle _{l_0}(\tau)= &
    \sum_{l=-\infty}^\infty \sum_{l'=-\infty}^\infty \frac{d_{2l}^* d_{2l'}}{2\pi} \\ 
    &\times \bra{l_0+2l} \cos^2(\theta) \ket{l_0+2l'}(\tau) \, .
    \label{2Dalignmentequation}
    \end{split}
\end{equation}
The integral
\begin{equation}
    \begin{split}
    &\bra{l_0+2l} \cos^2(\theta) \ket{l_0+2l'}(\tau) = \frac{1}{2} \delta_{l,l'} \\
    &+ \frac{e^{-i2\tau}}{4}\left( \delta_{l,l'+1} e^{i2( l_0 + 2l)\tau}+ \delta_{l+1,l'} e^{-i2(l_0 +2l)\tau}\right) \, 
    \end{split}
    \label{kronek} 
\end{equation}
in Eq.~\eqref{2Dalignmentequation} is only nonzero for specific combinations of $l$ and $l'$.

We insert the explicit form~\eqref{d2coeff} of the $d_{2l}$ coefficients into~\eqref{2Dalignmentequation}, and apply the discrete Kronecker deltas~\eqref{kronek} to the second sum in Eq.~\eqref{2Dalignmentequation} to obtain the alignment 
\begin{equation}
    \begin{split}
         \langle \cos^2(\theta)\rangle _{l_0}(\tau)=  \frac{1}{2} \sum_{l=-\infty}^\infty J_l\left(\frac{\mathrm{P}}{2}\right)^2
          &  \\
          - \frac{1}{2} \Re \bigg[i e^{i2(l_0+1)\tau} \sum_{l=-\infty}^\infty 
         e^{i4l\tau} & J_l\left(\frac{\mathrm{P}}{2}\right)  J_{l+1} \left(\frac{\mathrm{P}}{2}\right) \bigg]  \, .
     \end{split}
\end{equation}
Here both sums containing Bessel functions can be simplified through Graf's addition theorem~\cite{gradshteyn2014table}, that leads to the temporal alignment of a kicked eigenstate $\ket{l_0}$ in the form
\begin{equation}
    \langle \cos^2(\theta)\rangle _{l_0}(\tau)=  \frac{1}{2} + \frac{1}{2}\cos(2 l_0 \tau) J_1\left( \mathrm{P} \sin(2\tau)  \right) \, .
    \label{singlestatealign}
\end{equation}
Note that the time-dependent part of the alignment~\eqref{singlestatealign} factorizes into a pump-dependent term and a term depending on the initial state $l_0$. This is very useful, as an arbitrary-kicked state can be expressed as a superposition of kicked eigenstates. 

Now we consider a thermal ensemble, therefore the total alignment is given by the canonical sum
\begin{align}
    \langle \cos^2(\theta) \rangle(\tau) = \frac{1}{Z}\sum_{l_0=-\infty}^{\infty}  e^{-\frac{l_0^2}{2\sigma^2}}\langle \cos^2(\theta)\rangle _{l_0}(\tau) \, ,
    \label{canonical2D}
\end{align}
where we define the partition function
\begin{align}
    Z = \sum_{l_0 = -\infty}^\infty e^{-\frac{l_0^2}{2\sigma^2}} \,
    \label{2Dpartition}
\end{align}
and the dimensionless temperature 
\begin{align}
    \sigma = \hbar ^{-1} \sqrt{k_B T \mathcal{I}} \, 
    \label{defsigma}
\end{align}
with the moment of inertia $\mathcal{I}$ and the temperature $T$.

Inserting Eq.~\eqref{singlestatealign} into Eq.~\eqref{canonical2D} and after some straightforward algebra, we obtain the temporal alignment of a thermal ensemble of kicked 2D rotors as 
\begin{equation}
\begin{split}
     \langle \cos^2(\theta)\rangle (\tau) =  \frac{1}{2} + \frac{1}{2} &J_1\left( \mathrm{P} \sin(2\tau) \right) \\ 
     & \times\frac{\sum_{n=-\infty}^{\infty} e^{-2 \sigma^2 (\tau - n\pi)^2}}{\sum_{n=-\infty}^{\infty} e^{-2\sigma^2n^2 \pi^2 }} 
     \, .
\end{split}     
\label{2D-full}
\end{equation}
Eq.~\eqref{2D-full} reveals that the 2D alignment function is given by a pump-dependent oscillation multiplying a thermal envelope. Furthermore, the thermal envelope, in turn, is given by a normalized train of Gaussians. The width of the individual Gaussian decreases very quickly with rising temperature. For most applications the normalizing factor in Eq.~\eqref{2D-full} is very close to 1, which means that the overlap of the Gaussian functions is negligible and the envelope gives rise to individual revivals that are described by a Gaussian multiplying the pump-dependent oscillatory term.

\begin{figure}[t]
\includegraphics[width=0.98\linewidth]{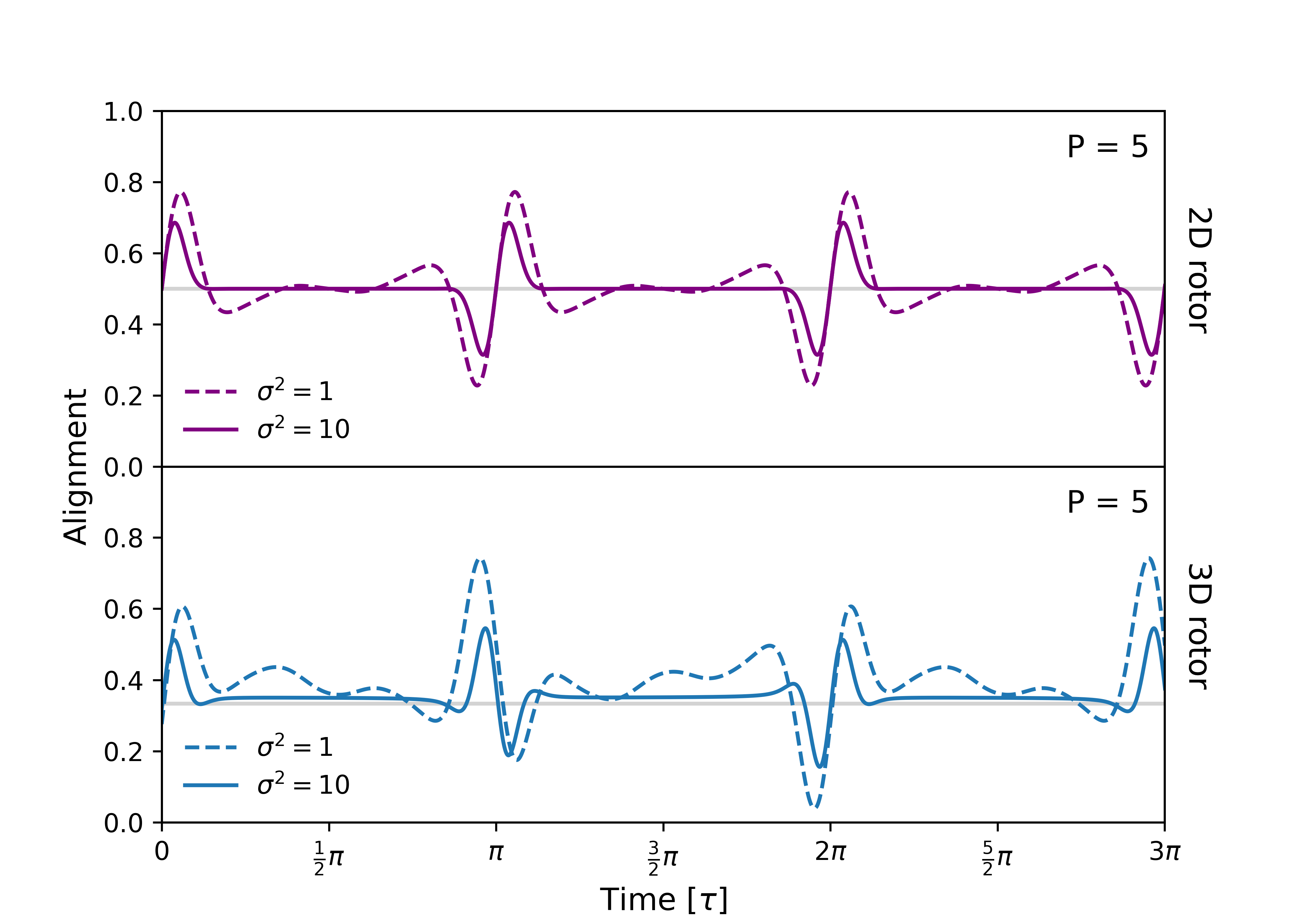}
\caption{Temporal alignment of the kicked 2D rotor~\eqref{2D-full} and 3D rotor~\eqref{3DAlignment} for initial temperature equivalents of $\sigma = 1$ and $ \sigma = \sqrt{10}$. The kick strength is P = 5. The gray line indicates the alignment of the respective initial thermally-populated system.}
\label{fig:2Dvs3D}
\end{figure}
The direct comparison of the 2D model~\eqref{2D-full} and full 3D model~\cite{Leibscher2004} (see Appendix~\ref{AppendixLeibscher}) for alignment at low and higher temperature is presented in Fig.~\ref{fig:2Dvs3D}. It shows that both models display isolated localized alignment oscillations, so-called rotational revivals. Furthermore, the shape of these revivals is very similar in the two models, with the revivals in 3D showing three main distinct deviations from the 2D case: (i) the revivals in 3D are narrower, (ii) they flip their parity periodically, and (iii) they have their average alignment value close to 1/3 instead of 1/2.

\section{3D alignment with a 2D rotor} \label{SectionHR}
We aim to construct a hybrid alignment model based on the simple 2D alignment~\eqref{2D-full} with physically-motivated adjustments in order to describe the 3D alignment~\eqref{3DAlignment}. To achieve this goal, we address the previously mentioned deviations between the two models, one at a time. 

We begin adjusting the 2D model to the 3D case by changing the field-free propagation energies of the 2D model in Eq.~\eqref{2DStatepropagation} to resemble its 3D counterpart:
\begin{align}
      c_{l}(\tau) = c_{l}  e^{-i\frac{l(l+1)}{2}\tau} \, .
      \label{HybridStatepropagation}
\end{align}

Inserting these states with their modified complex amplitude into the Eq.~\eqref{2Dalignmentequation} and following the derivation of the resulting time-dependent alignment analogously to the 2D case, we obtain a very similar equation with its periodicity altered due to an additional cosine factor. This cosine factor changes the parity of every second 2D revival, so it matches the sign of its 3D counterpart.

Next, we fix the difference in revival width which stems from the different number of degrees of freedom in the 2D and 3D models. We can account for this difference by multiply the temperature $T \sim \sigma^2$ in the 2D model Eqs.~\eqref{canonical2D} and~\eqref{2Dpartition} by $3/2$. 

Perhaps, the most noticeable consequence of the transition from 2D to 3D is that the mean value of the alignment for an isotropic system is $1/3$ in 3D instead of $1/2$ in 2D. This stems simply from the deviation in projection behavior between a circle and a sphere onto a single axis. Note that the mean value of the 3D alignment is, unlike the 2D model, not independent of the system parameters. The mean rises with the kick strength P, but also converges back to $1/3$ with increasing system temperature. This means that in most practical cases the mean is very close to $1/3$ as temperatures are non-cryogenic and the pump-pulse energy is limited by absorption due to plasma generation.

We therefore aim to transform the 2D alignment to have an average value of $1/3$. While there is a multitude of options, we choose to raise the alignment to the power $\mathrm{log}_2(3)$, ensuring that its values stay within the realistic interval $\{0,1\}$. 

Applying all the aforementioned adjustments results in our new model of the kicked hybrid-dimensional~(HD) rotor alignment:
\begin{equation}
\begin{split}
     \langle \cos^2(\theta)\rangle_\mathrm{HD}  (\tau) &=  \frac{1}{3}\bigg( 1 +  J_1{\left( \mathrm{P} \sin(2\tau) \right)} \cos(\tau) \\
     & \times 
     \frac{\sum_{n=-\infty}^{\infty} e^{-3 \sigma^2 (\tau - n\pi)^2}}{\sum_{n=-\infty}^{\infty} e^{-3\sigma^2n^2 \pi^2 }} 
     \bigg)^{\mathrm{log}_2(3)} \, .
     \label{HRmodel}
\end{split}
\end{equation}
Just like in case of the 2D model the denominator under the envelope in Eq.~\eqref{HRmodel} converges very quickly with rising temperature $\sigma$ and can safely be set to 1. 

As we demonstrate below, in Sec.~\ref{SectionResult}, the HD model catches the major characteristics of the 3D model in the vicinity of rotational revivals, even at low temperatures. As the areas between the revivals become flat rather quickly with rising temperatures, the overall accuracy of the model increases.

\section{Molecular model}\label{SectionMR}
Many molecules of interest exhibit central symmetry, most notably the homonuclear diatomic atmospheric gases N$_2$, O$_2$, H$_2$. Central symmetry in linear molecules implies that the Pauli principle has to hold under a rotation of $180^\circ$ in the rotor plane. We can formulate this condition according to Atkins~\cite{atkins2011molecular} as
\begin{align}
    (-1)^{l+I_g} = \pm 1 \, ,
    \label{Nuclear1}
\end{align}
where the sign is positive if the nuclei on either side of the molecule carry an integer spin and negative if the spin is half-integer. Furthermore, $l$ is the angular momentum of the molecule and $I_g$ is the total nuclear angular momentum given by the coupled angular momenta of the exchanged individual nuclei. This total angular nuclear momentum has strictly integer values for centrally-symmetric molecules, this means that for a given value of $I_g$ only every second value of angular momentum $l$ is allowed. 

To keep things simple, we focus on linear molecules comprised of two identical nuclei (homonuclear) only. For such molecules the total nuclear angular momentum is given by $I_g= \{ 0, .. , 2I_n\}$, where $I_n$ is the nuclear spin of one of the identical nuclei, restricting the values of $l$ through Eq.~\eqref{Nuclear1} to exclusively either even or odd for a given $I_g$. In general, this means that even and odd rotational states of homonuclear diatomic molecules are not equally populated even at thermal equilibrium. The canonical partition function, therefore, becomes 
\begin{align}
    Z = g_\mathrm{even}\sum_{l_0=\mathrm{even}} e^{-\frac{\tilde{E}_{l_0}}{\sigma^2}} + g_\mathrm{odd}\sum_{l_0=\mathrm{odd}} e^{-\frac{\tilde{E}_{l_0}}{\sigma^2}} \, ,
\end{align}
where $g_{l_0} = \{g_\mathrm{even},g_\mathrm{odd} \}$ is the statistical weight factor of even and odd states, respectively. The proportion of states $N_\mathrm{odd}$ with odd $l$ and states $N_\mathrm{even}$ with even $l$ for a diatomic molecule is given e.g.\ in~\cite{atkins2014atkins} and allows us to define the ratio of statistical weights $g_{l=\mathrm{even}} / g_{l=\mathrm{odd}} = N_\mathrm{even} / N_\mathrm{odd}$ as
\begin{align}
    \frac{g_{l=\mathrm{even}}}{g_{l=\mathrm{odd}}} = \left\{ 
    \begin{matrix}
    I_n/(I_n+1) & \text{ for half integer spin nuclei,}\\
    (I_n+1)/I_n & \text{ for integer spin nuclei.}
    \end{matrix}
    \right.
    \label{Nuclear2}
\end{align}

\begin{figure}[t]
    \centering
    \includegraphics[width=0.98\linewidth]{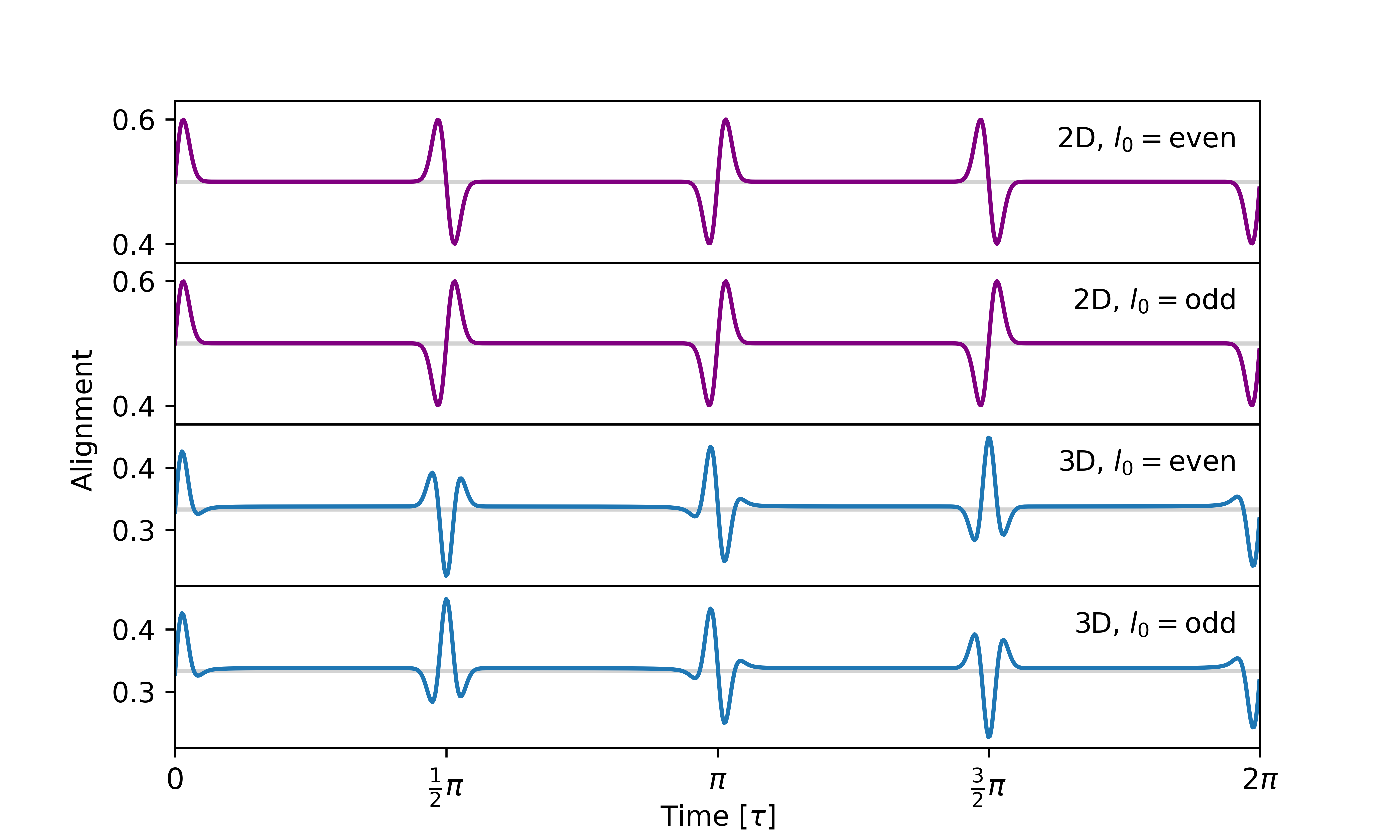}
    \caption{Temporal alignment of the 2D~\eqref{2D-full} and 3D~\eqref{3DAlignment} models for initial thermal distributions only considering either even or odd states. The kick strength $\mathrm{P}=7$ and temperature $\sigma=10$ are chosen. The gray line indicates the alignment of the respective initial thermally-populated system. }
    \label{fig:EvenOddPlots}
\end{figure}

To understand the physical manifestation of this separation into even and odd states, we demonstrate the temporal alignment function of a kicked rotor, treating the even and odd states separately, see Fig.~\ref{fig:EvenOddPlots}.

We can see that both even and odd alignment functions develop new revivals of similar amplitude at half integer multiples of $\pi$, so-called fractional revivals, while displaying practically identical full revivals. Because the fractional revivals of odd and even states have opposite signs, they cancel each other out and can not be observed in the previous rigid rotor models. In contrast to the 3D rigid rotor model, the mathematical emergence of fractional revivals in the 2D model comes naturally. 

If we recall Eq.~\eqref{singlestatealign} and Eq.~\eqref{canonical2D}, we can see that the shape of the thermal envelope in the molecular case is given by the Fourier series
\begin{align}
    \sum_{l_0=-\infty}^{\infty} \cos(2 l_0 \tau) e^{-\frac{l_0^2}{2 \sigma^2}} \, ,
\end{align}
which we can be expressed using the Fourier theorem as
\begin{equation}
\begin{split}
    \int_{-\infty}^\infty e^{-\frac{l^2}{2\sigma^2}}
    \left[ \sum_{n=-\infty}^\infty \delta(l-n) \right] \cos(2l\tau) dl &\\
    \sim e^{- 2 \sigma^2 \tau^2 } \circledast 
    \bigg[  \sum_{n=-\infty}^\infty \delta( \tau - n & \pi )\bigg] \, ,
\end{split}
\end{equation}
where the right-hand side is a convolution of a Dirac comb and the revival envelope. If we count only even or odd states, the periodicity of the Dirac comb in the frequency domain is doubled and therefore halved in the time domain.

However, a problem remains: while the 2D model shows promise in regards to the envelope of fractional revivals, it fails to capture the shape of the 3D fractional revivals as its power-dependent term $J_1(P \sin(2\tau))$ crosses zero exactly at half-integer multiples of $\pi$. The HD model shares this flaw as it shares the same factor.

While we have to accept that the HD model cannot describe the alignment of linear molecules in the vicinity of fractional revivals, we use our empirical observations to construct an approximate model based on HD model that we can extended to cover fractional revivals as well.

We do this by recognizing that all of the 3D revivals look as if they are pulses with the same envelope, but a phase shifted carrier. We can therefore obtain a good approximation for the fractional revivals if we can define a pseudo envelope and carrier within the 2D revival and shift the carrier of the 2D model revivals. 

If we use the existing separation into power-dependent oscillation and thermal envelope in Eq.~\eqref{2D-full}, we encounter two problems. First, the oscillation (a Bessel function of a sine) has no defined carrier frequency, therefore a phase shift of $\pi/2$ is not defined. Second, even if we were to find a suitable definition to approximate the oscillatory part of Eq.~\eqref{2D-full}, we can see in Fig.~\ref{fig:2Dvs3D} (and below in Fig.~\ref{3x3}) that for higher temperatures the observed frequency of oscillation is heavily-dependent of the system temperature $\sigma$. In fact, even for the full revivals the HD model misses the detail that the 3D revivals have additional small bumps at their edges. 

In order to define a carrier frequency that is higher than the oscillating Bessel function, while maintaining the slope of the revivals, we need to increase the width of the envelope. Our choice falls on factorization of the original temporal envelope $e^{-3\sigma^2\tau^2}$ into two equal parts, keeping one of the two envelopes (whose width is increased by $\sqrt{2}$) as a multiplying envelope, while combining the second with the oscillatory term and approximating it with a sine function. For the main revivals we therefore look for an approximation of the form
\begin{align}
    J_1\left\{ P \sin(2\tau)\right\}e^{-3\sigma^2\tau^2} \approx A \sin( B \tau ) e^{-\frac{3}{2}\sigma^2\tau^2} \, .
    \label{CarrierEnvelopeApprox}
\end{align}
By expanding both sides around zero up to third order and matching the coefficients, we get
\begin{equation}
\begin{split}
    A = & \mathrm{P}/\sqrt{3\mathrm{P}^2+9\sigma^2+4} \, , \\
    B = & \sqrt{3\mathrm{P}^2+9\sigma^2+4} \, .
\end{split}
\label{CoefficientsHightemp}
\end{equation}

Now that we have defined an envelope and a carrier frequency, we can simply shift of the carrier phase by $\pi /2$ to approximate fractional revivals. The molecular rotor (MR) model alignment of a thermally-populated ensemble of molecules can be approximated by summing up full and fractional revivals as
\begin{align}
    \begin{split}
    &\langle \cos^2(\theta) \rangle_\mathrm{MR} (\tau) = \frac{1}{3}
    \bigg[
    1 + A\sum_{n=-\infty}^{\infty} (-1)^n  \\ &\times \Bigg\{ \sin\left( B (\tau - n\pi) \right) e^{-\frac{3}{2} \sigma^2 (\tau - n\pi)^2} 
    + \frac{g_\mathrm{even}-g_\mathrm{odd}} {g_\mathrm{even}+g_\mathrm{odd}} 
      \\ & \times\cos \left( B (\tau - (n + 1/2)\pi) \right) e^{-\frac{3}{2} \sigma^2 \left(\tau - (n + 1/2)\pi \right) ^2}  
      \Bigg\} \Bigg] ^{\log_2(3)} \, .
    \end{split}
    \label{MRmodel}
\end{align}

We demonstrate below in the following Sec.~\ref{SectionResult} that the MR model~\eqref{MRmodel} is capable of describing both full and fractional revival features for a large range of system parameters.

\section{Model Comparison}\label{SectionResult} \label{SectionResult}
\begin{figure*}[t]
\includegraphics[width=0.8\linewidth]{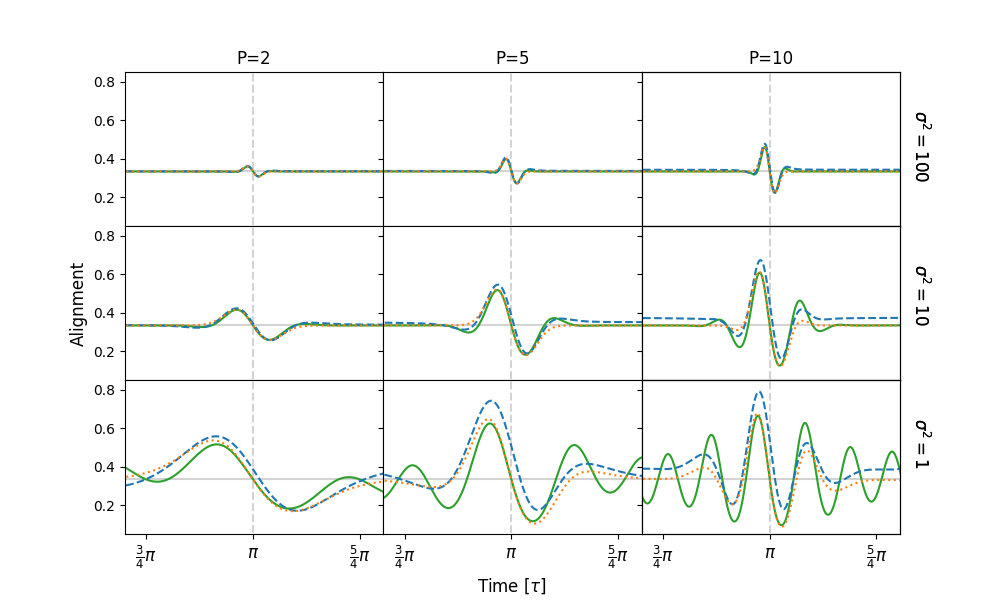}
\caption{Temporal alignment of the HD model (dotted orange), the MR model (solid green) and the exact 3D model (dashed blue) in the vicinity of the full revival ($\tau = \pi$) for all combinations of temperatures $\sigma^2 = 1, 10, 100$ and kick strengths $\text{P} = 2, 5 ,10$.}
\label{3x3}
\end{figure*}

We compare the performance of our MR model~\eqref{MRmodel} and HD model~\eqref{HRmodel} for alignment revivals with the one calculated by the exact 3D model~\cite{Leibscher2004} (see Appendix~\ref{AppendixLeibscher}). This comparison of the models is illustrated in Fig.~\ref{3x3} for various gas temperatures and laser kick strengths.

Strong agreement of all three models can be seen for high temperature $\sigma^2 =100$ (corresponding to \SI{600}{K} for N$_2$) and, with exception for the high power case of $\mathrm{P}=10$, for moderate temperature $\sigma^2 =10$ (corresponding to \SI{60}{K} for N$_2$). For low temperature $\sigma^2 = 1$ (corresponding to \SI{6}{K} for N$_2$) the MR model loses accuracy rapidly outside of the revival center while the HD model is still able to capture the main oscillatory features of the 3D model. 

The breakdown of the MR model for low temperatures and high kick powers is expected as it is based on a Taylor expansion around the revival center. We estimate the condition for the applicability of the MR model through $\mathrm{P} < \sqrt{2}\sigma$, see the derivation in Appendix~\ref{AppendixApplicability}.

Next, we look at the ability of our models to capture the features of fractional revivals. We exclude the HD model from this comparison due to its previously discussed inability to recreate the cosine shaped fractional revivals.
\begin{figure}[h]
    \centering
    \includegraphics[width = 0.98\linewidth]{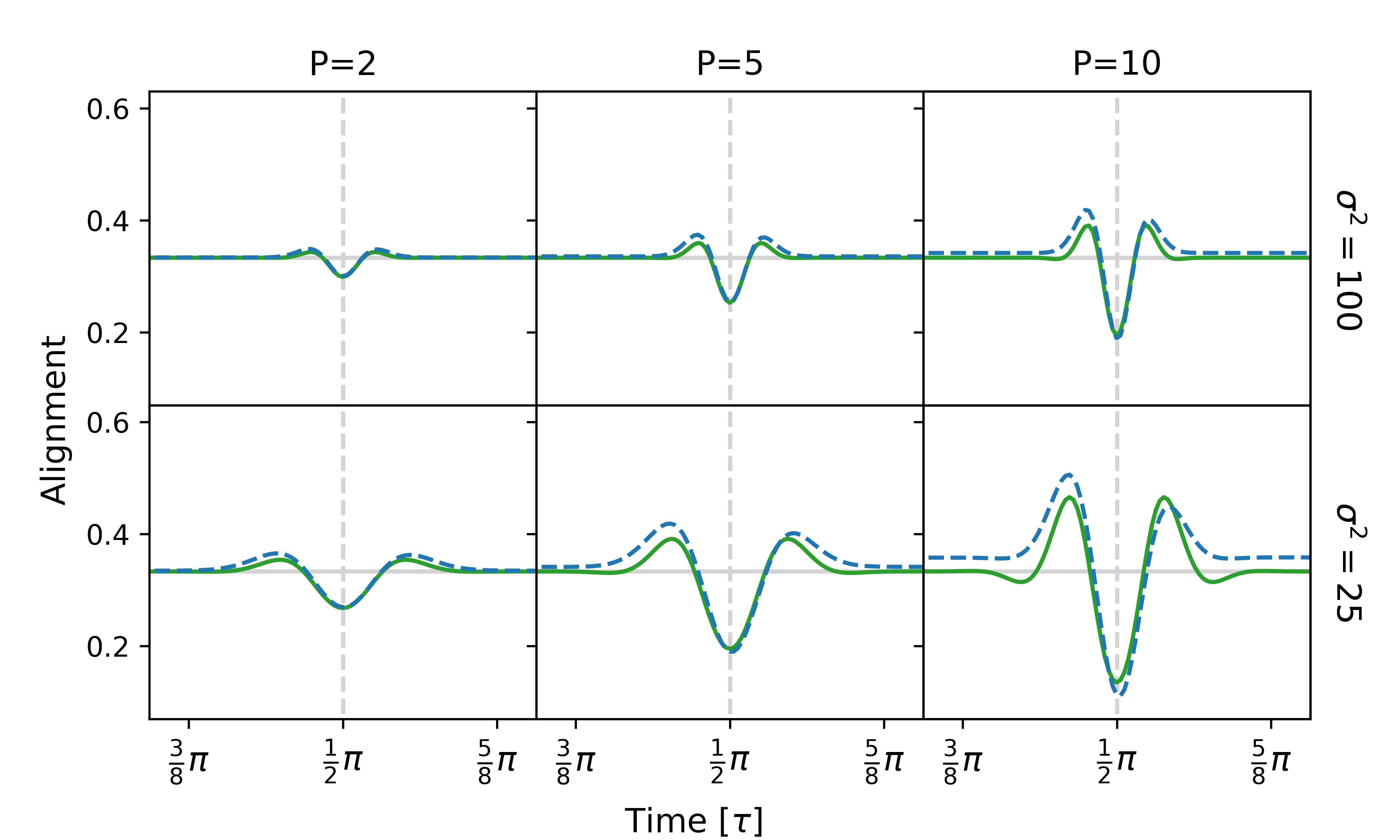}
    \caption{Fractional alignment revivals for different $\mathrm{P}$ and $\sigma^2$. Results of the MR model~\eqref{MRmodel} (solid green) and 3D model~\eqref{3DAlignment} (dashed blue), for initial thermal distributions only considering even states ($g_{\text{odd}} = 0$). The gray line corresponds to the 3D isotropic alignment value of 1/3.}
    \label{fig:fractional}
\end{figure}

Fig.~\ref{fig:fractional} depicts the performance of the MR model in approximating fractional rotational revivals for representative set of system parameters. Lower temperatures are not displayed as the requirements for the validity of the MR model are not met. While the MR model slightly underestimates the alignment at the revival edges, it shows good agreement with the 3D model for moderate kick strengths and higher temperatures.

\section{Alignment of atmospheric molecules} \label{RealMolecules}
Here we provide a practical demonstration of the MR model by applying it to the main atmospheric molecules N$_2$, O$_2$ and CO$_2$ under experimental conditions commonly found in hollow-core fibers~\cite{haddad2018molecular,nagy2021femtosecond, nagy2021high}. The three molecules are chosen due to their natural abundance and differences in anisotropic polarizability, moment of inertia as well as nuclear spins. 

We assume all molecules to be thermally-populated at \SI{300}{K} and kicked by a pulse energetically equivalent to a square pulse with the duration \SI{100}{fs} and an intensity of $\SI{3.3e13}{W/cm^2}$. This results in different dimensionless system parameters for each molecule, which are shown in Table~\ref{tab:my_label}.
\begin{table}[h!]
    \centering
    \begin{tabular}{c|c|c|c|c}
         & P & $\sigma$ & $I_n$ & $\mathcal{I}$ \\ \hline
       N$_2$  & 4.5 & 7.2 & 1 & 1.40 \\ \hline
       O$_2$  & 7.1 & 8.5 & 0 & 1.94\\ \hline
       CO$_2$ & 15.3 & 16.4 & 0 & 7.15
    \end{tabular}
    \caption{Dimensionless molecular parameters~\eqref{defP} and~\eqref{defsigma} for the three different atmospheric molecules under the same experimental conditions. Polarizabilities $\Delta \alpha$ taken from~\cite{gough1996analysis, spelsberg1994static, qin2014probing}, moments of inertia $\mathcal{I}$ in units of $\SI{e-46}{Kg m^2}$ and nuclear spin $I_n$ are taken from~\cite{huber2013molecular}.}
    \label{tab:my_label}
\end{table}

\begin{figure}[h!]
\includegraphics[width=0.98\linewidth]{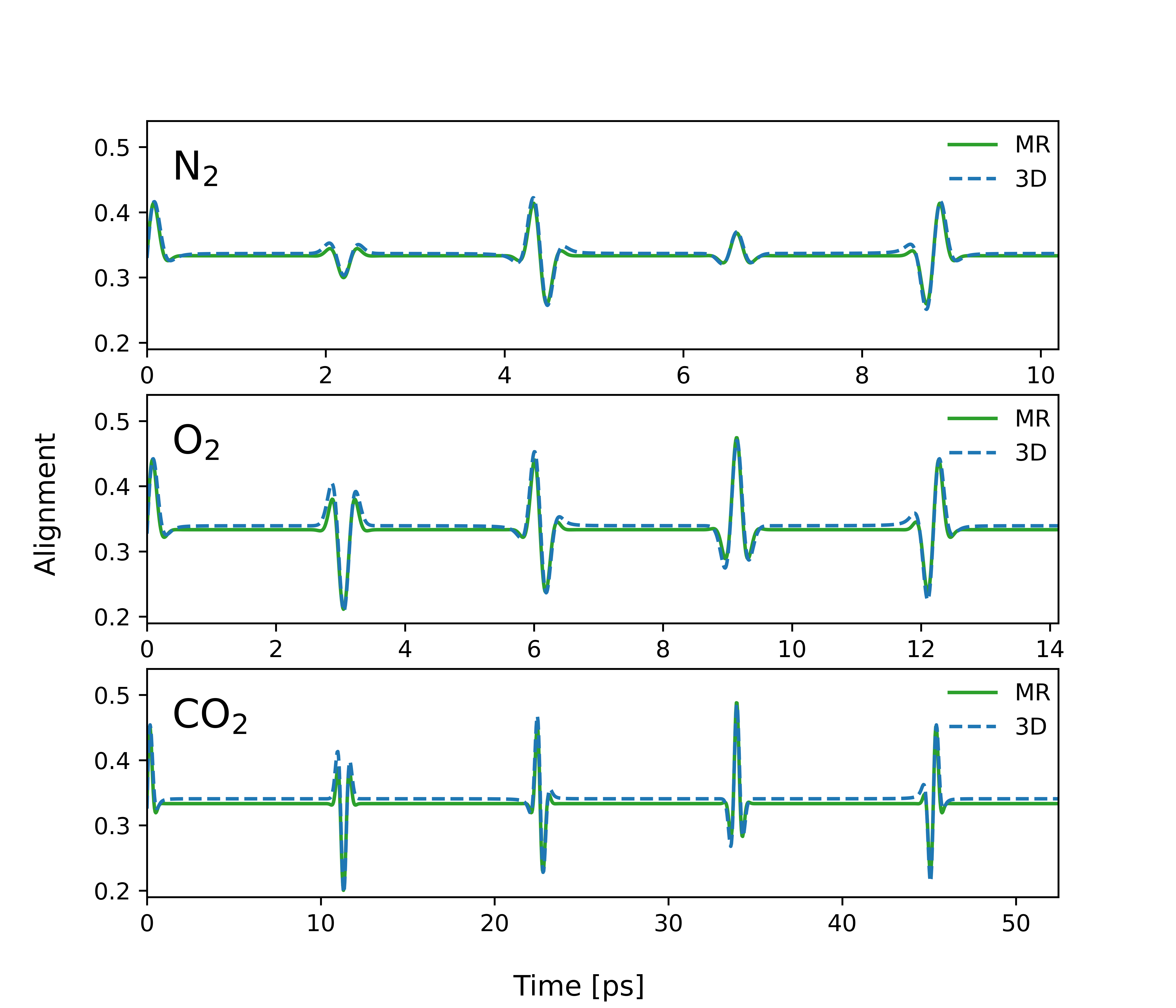}
\caption{Temporal alignment of kicked N$_2$, O$_2$ and CO$_2$ molecular ensembles according to the MR model~\eqref{MRmodel} and the 3D model~\eqref{3DAlignment}. In all cases the molecules are assumed to be thermally-populated at \SI{300}{K} and kicked by an energy equivalent of a \SI{100}{fs} square pulse with intensity $\SI{3.3e13}{W/cm^2}$.} 
\label{MolecularComparison}
\end{figure}

In Fig.~\ref{MolecularComparison} we can see that the different kicked molecules display distinctly different alignment dynamics.
Comparing the alignment of N$_2$ and O$_2$, we see that the timescales of revival occurrence are fairly similar due to the similar moment of inertia of the two molecules differing by only 40\%. The main qualitative difference is the amplitude of the fractional revivals, which appears suppressed in N$_2$ due to its nonzero nuclear spin. The alignment timescales of O$_2$ and CO$_2$, on the other hand, differ significantly due to their moment of inertia differing by almost 400\%. The qualitative alignment behaviors for the two molecules are, however, very similar due to the shared nuclear spin of zero, resulting in the fractional revivals not being suppressed.

In summary for the model comparison, Fig.~\ref{MolecularComparison} shows that the MR model is able to capture the alignment characteristics of all three molecules very successfully, demonstrating that it is an effective tool for describing the impulsive alignment of various linear molecules under realistic conditions.

\section{Conclusions}
In this paper we have developed a simple alignment model for thermal ensembles of linear rotors and molecules kicked by a short laser pulse, which is based on 2D rotor alignment. We have demonstrated its performance for a wide range of parameters by comparing its ability to describe the shape of rotational revivals against the exact 3D solution~\cite{Leibscher2004}. 

We have found that our HD, or hybrid-dimensional, rotor model is able to describe the alignment features of 3D rotors accurately while being mathematically significantly simpler than the full 3D model~\cite{Leibscher2004}. Furthermore, the HD model is sensitive to the individual system parameters in a clear way, with the kick strength providing a carrier oscillation in the alignment which is multiplied by a thermal envelope that gives rise to individual rotational revivals.

Armed with this simple HD rotor model, we have adapted it further to describe the fractional revivals that occur in molecules exhibiting central symmetry. The resulting MR, or molecular rotor, model has shown to be accurate for alignment calculations of thermally-populated molecular ensembles in a wide range of parameters that are standard in many experimental scenarios. The results obtained within our MR model only diverge from the exact solutions for cold molecules that are kicked very hard.

For demonstration purposes we have applied the MR model to alignment calculations of the three atmospheric gases N$_2$, O$_2$ and CO$_2$ under experimental conditions commonplace in hollow-core fibers and found excellent agreement between the 3D and MR models.

To summarize, the analytical alignment MR model we have introduced offers a very simple yet accurate model for the time-dependent alignment of laser-kicked thermal ensembles of molecular rigid rotors. We believe that its simplicity will prove useful beyond analytical and numerical calculations as it establishes an intuitive correspondence between system parameters, such as pulse energy, gas temperature and molecular species, and their influence on the shape of the resulting alignment behavior. 

Moreover, we have already demonstrated~\cite{Lohr2023} that the MR model enables an easy access for the study of propagation effects such as compression, frequency up-conversion and amplification of a laser pulse in rotationally-kicked molecular gases.

\newpage
\appendix
\begin{widetext}
\renewcommand{\theequation}{\Alph{section}.\arabic{equation}}
\numberwithin{equation}{section}
\section{Alignment of the 3D kicked rigid rotor}\label{AppendixLeibscher}
The alignment of a 3D thermal ensemble of rigid rotors is given by Leibscher~\cite{Leibscher2004} as
\begin{align}
    \langle\cos^2(\theta)\rangle_{3D}(\tau) = \frac{1}{Z}\sum_{l_0=0}^{\infty} \sum_{m_0=-l_0}^{l_0} e^{-\frac{l_0(l_0+1)}{2 \sigma^2}}\langle\cos^2(\theta)\rangle_{l_0,m_0} \, ,
    \label{3DAlignment}
\end{align}
where
\begin{align*}
    \langle \cos^2(\theta)\rangle_{l_0,m_0}(\tau) &= \frac{1}{3}\sum_{l'=0}^\infty \frac{(2l'+3)(2l'-1)+[l'(l'+1)-3m_0^2]}{(2l'+3)(2l'-1)}|a(l_0,m_0,l')|^2   \nonumber \\& +2 \Re \left[ \sum_{l'=0}^\infty a(l_0,m_0,l')a(l_0,m_0,l'+2)^*\frac{1}{2l'+3}\sqrt{\frac{[(l'+2)^2-m_0^2][(l'+1)^2-m_0^2]}{(2l'+5)(2l'+1)}}e^{i(2l'+3)\tau} \right]
\end{align*}
with
\begin{align*}
    a(l_0,m_0,l')=(-1)^{m_0}\sum_{l=0}^{\infty} \sqrt{\frac{(4l+1)(2l_0+1)(2l'+1)}{4\pi}}\begin{pmatrix}
2l & l_0 & l' \\
0 & m_0 & -m_0
\end{pmatrix}\begin{pmatrix}
2l & l_0 & l' \\
0 & 0 & 0
\end{pmatrix}c_{l} 
\end{align*}
and
\begin{align*}
    c_l = \sqrt{\pi(4l+1)}(i \mathrm{P})^l \frac{\Gamma(l+1/2)}{\Gamma(2l+3/2)} \text{}_1F_1(l+\frac{1}{2},2l+\frac{2}{3}, i \mathrm{P}) \, ,
\end{align*}
where $\Gamma$ and $\text{}_1F_1$ are the gamma and Kummer's confluent hypergeometric functions, respectively. 

\end{widetext}

\section{Validity of the MR model}\label{AppendixApplicability}
To estimate the realm of validity of the MR model~\eqref{MRmodel}, we set two constraints. First, for all the revivals to be considered separately, their envelopes cannot overlap. Second, if the Bessel function that is being approximated does not get flattened fast enough by the envelope, the sinusoidal approximation will be inaccurate. 

We define the revivals to be separated when their envelopes have less than 1\% of their peak value at half the distance between revivals. As the sinusoidal approximation has a wider envelope than the HD model, we use its envelope as an upper bound measure. With MR model having full and fractional revivals, we write this condition as 
\begin{align}
    e^{-\frac{3 \sigma^2}{2} \left(\frac{\pi}{4} \right)^2} \leq 0.01 \, ,
\end{align}
which is equivalent to
\begin{align}
    \sigma \geq \sqrt{\frac{32}{3\pi^2} \ln(100)} \approx 2.3 \, .
    \label{Condiditon1CE}
\end{align}

The carrier and envelope approximation states
\begin{align}
    J_1\left( P \sin(2\tau)\right) e^{-\frac{3}{2}\sigma^2\tau^2} \approx A \sin( B \tau ) \, .
    \label{semicondi1}
\end{align}
This approximation is clearly not general, as one side decays while the other does not. This is not a problem if the envelope multiplying both decays fast enough. We can estimate this condition by requiring the second lobe of the sinusoidal carrier to be suppressed by more than 95\%.
The time delay $\Delta \tau$ of the second lobe of the carrier is given by
\begin{align}
    \Delta \tau = \frac{3 \pi}{2 B} \, .
\end{align}
Therefore, we can formulate the aforementioned condition as
\begin{align}
    e^{-\frac{3}{2}\sigma^2 (\Delta \tau )^2} \leq 0.05
\end{align}
or equivalently
\begin{align}
    \mathrm{P}^2  \leq 3 \left[ \frac{9}{8}\frac{\pi^2}{\ln(20)} - 3\right]\sigma^2 - \frac{4}{3} \, .
    \label{semicond3}
\end{align}
Since $\sigma \geq 2.3$, the second term on the right-hand side in Eq.~\eqref{semicond3} can, in general, be neglected and we can estimate an upper bound as
\begin{align}
    \mathrm{P}\leq \sqrt{2}\sigma \, .
\end{align}

\section*{Acknowledgments}
A.L. acknowledges support of the ERC-2021-AdG project ULISSES, grant No.~101054696. 
M.I. acknowledges support of the H2020 European Research Council Optologic grant (899794). 
This work was supported by the Leibniz Gemeinschaft (SAW-2021-MBI, project No.: K380/2021).

\bibliography{lit}

\end{document}